# Micropolarity Effects on the Bickley-Plane-Laminar-Jet

Abuzar A. Siddiqui

Department of Basic Sciences, Bahauddin Zakariya University, Multan, Pakistan.

**Abstract:** In this study, it was formulated the boundary-value-problem (BVP), comprising partial differential equations (PDEs), of steady flow for plane, laminar jet of a micropolar fluid. A new similarity transformation/solution was derived which is valid not only for the Newtonian fluids but also for the micropolar fluids. Obviously, this transformation will be transformed the PDEs into the ordinary differential equations (ODEs). These ODEs were solved numerically by the finite difference method. The obtained results were compared with existing results [1, 12] for the Newtonian fluids. The comparison was favourable. As the aciculate particles in a micropolar fluid can rotate without translation, the micropolarity effects must have influence on fluid-speed, microrotation, stresses, couple stresses and discharge. This influence was highlighted in the present study. If viscosity coupling parameter $K_1$ (being the measure of micropolarity) increases then microrotation, fluid-flux, stresses and couple stresses $(m_{13} \& m_{31})$ intensify in the vicinity of the jet along $y$-direction. The fluid-flux, $Q \propto x$, for a fixed value of $K_1$ and for the micropolar as well as Newtonian fluids. Moreover, the normal stresses are related mutually as $\sigma_{11} = -\sigma_{22}$. In addition, the stress and the couple stress tensors are non-symmetric but the couple stress tensor will be symmetric and skew-symmetric, respectively, if $\beta \mp \gamma = 0$.

## 1. Introduction

There is a type of fluids comprising aciculate rigid/semi-rigid particles which can rotate (without translation) about the axes passing through their centroids, are named as micropolar fluids. In 1966, Eringen [2] derived their governing and constitutive equations. These fluids sustain asymmetric stress tensor in addition to couple stress tensor (which occurs owing to the spinning of the colloids/aciculate particles). In addition, these fluids have six degree of freedom---three more than the Newtonian fluids. Micropolar fluids are exemplified by blood [3-4], polymers and polymeric suspension [5], rigid-rod epoxies [6], colloidal suspension and liquid crystals [7]. These fluids are being used on lab-on-a-chip [8-9] nowadays. These fluids have very healthy history and a lot of work has been done. We shall not survey the history here but we shall refer [2, 7, 10-11, 21-22] for the reader who has interest to know their history.

On motivating the technoscientifical and industrial importance of micropolar fluids, we embarked the program to probe the micropolarity effects on the laminar plane jet. A laminar plane jet is analogous to the point/line source of fluid, that can emit fluid from a long but narrow slit and mixes with the surrounding fluid [12]. It gained great attention not only in the previous century like [1, 12] but also it is burning issue of the day in the current century [13-14]. It is due to fact that, it has tremendous and marvelous applications in industry and engineering. For example: in setting up the ink-jet printers [15-17], in cooling hot material like steel plates at ROT [13], in cooling of combustion engines and electronic microchips [14], in evaporation of refrigerant-oil mixture [18], in Erosion threshold of a liquid immersed granular bed [19] and much more in [20].

The plan of this paper is: Section 2 comprises mathematical formulation of the boundary value problem for under-consideration flow and similarity transformation; the computational procedure will be presented in section 3 while section 4 contains brief review of the Bickley-Schlichting (BS) Solution for the Newtonian Fluids. The results (in graphical form) in addition to the discussion on it, are given in section 5 whereas conclusions are made briefly in section 6.

## 2. Mathematical Formulation

Consider a steady "The Plane Jet" flow [1] having width $w$, of viscous and incompressible micropolar fluid. Its governing equations, originally derived by [7], in dimensional form are

$$\nabla' \cdot \mathbf{V}' = 0, \tag{1}$$

$$-(\mu+\chi)\nabla' \times (\nabla' \times \mathbf{V}') + \chi \nabla' \times \mathbf{N}' - \nabla'\Phi + \rho\mathbf{B} = \rho(\mathbf{V}' \cdot \nabla')\mathbf{V}', \tag{2}$$

$$(\alpha+\beta+\gamma)\nabla'(\nabla' \cdot \mathbf{N}') - \gamma\nabla' \times (\nabla' \times \mathbf{N}') + \chi(\nabla' \times \mathbf{V}' - 2\mathbf{N}') + \rho\mathbf{C} = \rho j_o (\mathbf{V}' \cdot \nabla')\mathbf{N}', \tag{3}$$

with respect to the following constitutive equations [21-22];

$$\underline{\underline{\sigma'}} = \lambda \underline{\underline{I}} \nabla' \cdot \mathbf{V'} + \left(\mu + \frac{\chi}{2}\right)\left[\nabla'\mathbf{V'} + (\nabla'\mathbf{V'})^T\right] - \frac{\chi}{2}\underline{\underline{I}} \times (\nabla' \times \mathbf{V'}) + \chi \underline{\underline{I}} \times \mathbf{N'},$$

(4)

$$\underline{\underline{m'}} = \alpha \underline{\underline{I}} \nabla' \cdot \mathbf{N'} + \beta \nabla' \mathbf{N'} + \gamma (\nabla' \mathbf{N'})^T. \tag{5}$$

Here $\underline{\underline{I}} = \nabla'\mathbf{r'}$ is the idempotent and superscript T stands for transpose. In addition, $\mathbf{V'}$ and $\mathbf{N'}$, respectively are the dimensional fluid velocity and the microrotation of collides; $\rho$ and $j_o$ are, respectively, the fluid density and the microinertia; $\Phi$, $\mathbf{B}$, and $\mathbf{C}$ are, respectively, the hydrostatic pressure, the body-force and the body-couple; and $\alpha$, $\beta$ and $\gamma$ are three spin-gradient viscosity coefficients whereas $\mu$ and $\chi$ signify for the spin viscosity coefficients such that $2\mu + \chi \geq 0$ where $\chi \geq 0$ [21].

As far as the plane jet is concerned, it is defined as earlier: it is a long narrow orifice/slit of fluid in a stationary fluid that ejects the fluid continuously, freely and steadily in two dimensional [1]. The (x, y)-coordinates system will be suitable such that the origin is lying on the orifice-centre whereas the x-axis is taken along the axis of symmetry of the plane jet, as shown in figure 1.

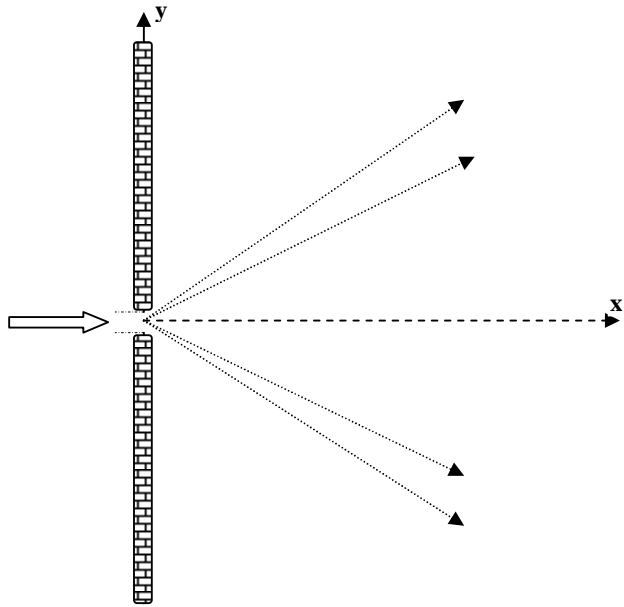

**Fig. 1** The Schematic of the flow

Moreover, this study was made under the following assumptions:

- The flow is considered as two dimensional so the velocity and microrotation can be expressed as $\mathbf{V}' = [u'(x', y'), v'(x', y'), 0]$ and $\mathbf{N}' = [0, 0, S'(x', y')]$ respectively.
- The (emerging) jet spreads outward in the downstream direction on account of the effect of the friction.
- The ambient fluid is not influenced by any external forces like the body-force and the body-couple.
- The (hydrostatic) pressure is assumed to be constant and is independent of coordinate axes. Whence it can be concluded that the momentum along x-direction $M_x$ will remain constant i.e. $M_x = \rho \int_{-\infty}^{\infty} u'^2 dy' =$ constant [12].
- The velocity (u')-profile may be analogous to that of flow past a flat plate at zero incidence [12].
- The velocity-component $u'$ will be function of $y'/w$ [12], where w denotes for the width of the jet.
- The presumed Prandle boundary layer theory for a Newtonian fluid will be preserved for under consideration micropolar fluid.

Equs. (1) – (3), in the light of above assumptions, will take the form as:

$$\frac{\partial u'}{\partial x'} + \frac{\partial v'}{\partial y'} = 0 \tag{6}$$

$$u'\frac{\partial u'}{\partial x'} + v'\frac{\partial u'}{\partial y'} = v_m \frac{\partial^2 u'}{\partial y'^2} + \frac{\chi}{\rho} S' \tag{7}$$

$$\rho j_o v' \frac{\partial S'}{\partial y'} + \chi \frac{\partial u'}{\partial y'} = \gamma \frac{\partial^2 S'}{\partial y'^2} - 2\chi S' \tag{8}$$

Where $v_m = (\mu + \chi)/\rho$. The boundary conditions will be

$$v = 0, \quad \frac{\partial u'}{\partial y'} = 0, \quad u' = \frac{v_m}{w}, \quad S' = 0 \qquad \text{at} \quad y' = 0 \tag{9}$$

$$u' \to 0 \text{ and } S' = 0 \qquad \text{as} \quad y' \to \infty \tag{10}$$

by virtue of no-slip-no-spin and matching conditions.

On applying the dimension analysis on above boundary value problem (6) - (8), we get

$$u\frac{\partial u}{\partial x} + v\frac{\partial u}{\partial y} = \frac{\partial^2 u}{\partial y^2} + K_1 S \tag{11}$$

$$v\frac{\partial S}{\partial y} = K_2 \frac{\partial^2 S}{\partial y^2} - K_3\left(2S + \frac{\partial u}{\partial y}\right) \tag{12}$$

where $(u, v) = w v_m^{-1} (u', v')$, $(x, y) = w^{-1}(x', y')$ and $S = w^2 v_m^{-1} S'$ such that $K_1 = \chi/(\mu + \chi)$, $K_2 = \gamma/(\mu + \chi) j_o$ and $K_3 = (w^2 K_1)/j_o$. In addition, the dimensional stress $\underline{\underline{\sigma}}'$ and couple stress tensors $\underline{\underline{m}}'$ are related with their corresponding dimensionless form as: $\underline{\underline{\sigma}} = \rho w^2 (\mu + \chi)^{-2} \underline{\underline{\sigma}}'$ and $\underline{\underline{m}} = w^3 (\gamma v_m)^{-1} \underline{\underline{m}}'$.

Now, let us introduce similarity solution, in order to transform partial differential equations given in Eqs. (11) - (12) into ordinary differential equations, as:

$$u = x \frac{df(y)}{dy}, \quad v = -f(y) \text{ and } S = xg(y) \tag{13}$$

On using Eq. (13) into Eqs. (11) - (12), we yield

$$\frac{d^3 f}{dy^3} + f \frac{d^2 f}{dy^2} - \left(\frac{df}{dy}\right)^2 + K_1 \frac{dg}{dy} = 0 \tag{14}$$

$$K_2 \frac{d^2 g}{dy^2} + f \frac{dg}{dy} - K_3 \left(2g + \frac{d^2 f}{dy^2}\right) = 0 \tag{15}$$

The boundary conditions will be deformed as:

$$f = 0, \; g = 0, \; \frac{d^2 f}{dy^2} = 0 \text{ and } \frac{df}{dy} = 1 \qquad \text{at } y = 0 \tag{16a}$$

$$g \to 0 \text{ and } \frac{df}{dy} \to 0 \qquad \text{as } y \to \infty \tag{16b}$$

The boundary value problem (BVP) given above comprises coupled, non-linear ordinary differential equations and cannot be solved analytically. Therefore, it will be solved by the numerical scheme whose description is given in next section.

### 3. Computational Procedure

Eqs. (14) - (15) may be written as:

$$\frac{d^2 \phi}{dy^2} + f \frac{d\phi}{dy} - \phi^2 + K_1 g = 0 \tag{17}$$

$$K_2 \frac{d^2 g}{dy^2} + f \frac{dg}{dy} - K_3 \left(2g + \frac{d\phi}{dy}\right) \tag{18}$$

where $\phi = \frac{df}{dy}$. $\tag{19}$

The central finite difference approximations are applied to the derivatives involved in Eqs. (17) - (18) to get a set of two finite difference equations. These are further solved by the Successive-Over-Relaxation method [23] whereas Eq. (19) will be integrated by the Simpson's rule [24].

## 4. Bickley-Schlichting (BS) Solution and the Newtonian Fluids

Before discussing the results, let us review and reproduce the existing solution for a Newtonian fluid, which was developed by Bickley-Schlichting [1, 12] and let us name it Bickley-Schlichting (BS) solution.

If we put $K_1 = 0$ in Eq. (3), we get the boundary layer equation for the Newtonian fluid

$$u'\frac{\partial u'}{\partial x'} + v'\frac{\partial u'}{\partial y'} = \frac{\mu}{\rho}\frac{\partial^2 u'}{\partial y'^2} \tag{20}$$

Bickley-Schlichting (BS) developed the following similarity solution [1] and [12]:

$$u' = \frac{1}{3}x^{-1/3}\frac{df}{d\eta} \tag{21}$$

$$v' = -\frac{1}{3}\left(\frac{\mu}{\rho}\right)^{1/2}x^{-2/3}\left(f - 2\eta\frac{df}{d\eta}\right) \tag{22}$$

where $\eta = \frac{1}{3}\left(\frac{\mu}{\rho}\right)^{-1/2}x^{-2/3}y$.

On using (21) - (22) in Eq. (20), it can easily be found:

$$\frac{d^3 f}{d\eta^3} + f\frac{d^2 f}{d\eta^2} + \left(\frac{df}{d\eta}\right)^2 = 0 \tag{23}$$

This equation was solved analytically with subject to the aforementioned boundary conditions by Bickley-Schlichting as: [1, 12]

$$u' = 0.4543\left(\frac{k^2\rho}{\mu x}\right)^{1/3}\left[1 - \tanh^2(\xi)\right] \tag{24}$$

where $\xi = 0.2752\left(\frac{k\rho^2}{\mu^2}\right)^{1/3}x^{-2/3}y = \alpha_1 y$ such that $\rho k = M_x$.

## 5. Results and Discussions

All the calculation are carried out for fixed material constants $\mu = 3 \times 10^{-2}$ Pa.s, $\rho = 1.2 \times 10^3$ Kg.m$^{-3}$ [21]. The couple stress parameter $\beta \in [-\gamma, \gamma]$ and viscosity coupling parameter $K_1 \in [0,1)$ are used [21].

The effect of micropolarity is highlighted on laminar plane jet for physical parameters of fluid like fluid-speed, fluid-flux, fluid-stresses and fluid-couple stresses graphically in this section.

The fluid-velocity has two non-zero components horizontal-fluid-speed $u$ and transverse-fluid-speed $v$. These are displayed in figures 2 & 3; figure 2 depicts for variation of $u$ at $x=1$ with $y$ for three values of $K_1$ namely $K_1=0$ (for the Newtonian fluids), $K_1=0.01$ and $K_1=0.005$. It can be observed that, horizontal-fluid-speed u decreases with increasing y for the Newtonian as well as for the micropolar fluids. In addition $u \to 0$ as $y \to \infty$, owing to the fact that the fluid is stationary far from the jet, which is in justification with boundary condition (10) for both the Newtonian and micropolar fluids. The distinguished feature of micropolar is that, the reverse flow exists if we increase $K_1$ as $u<0$ for the micropolar case e.g. for $K_1=0.005$ (as presented in figure 2), while $u$ remains positive for the Newtonian fluids for all values of x and y. Consequently the boundary layer separation occurs in the vicinity of jet. An observation which is not displayed here is that boundary layer separation points grow as $K_1$ increases. Moreover, figure 2 also shows the comparison of the present results for the Newtonian fluids ($K_1=0$) with the existing results [1, 12]. Both the results compare well for moderate and high values of $y$ but it differs slightly for low values of $y$. However, the decreasing trend of $u$ with $y$ of both the results is concurrent.

Unlike the horizontal component of velocity, the transverse-fluid-speed $v$ is independent of $x$, which is in accordance with the similarity solution given in (13). This transverse fluid speed $v$ decreases as $y$ increases for all $x$ but for a Newtonian fluid and very low value of $K_1$ as shown in figure 3. In addition, $v$ increases in the vicinity of jet for high value of $K_1$ near jet. However, the common feature for both the fluids (Newtonian as well as micropolar) v remains constant as $y \to \infty$ for all $x$ and $K_1$.

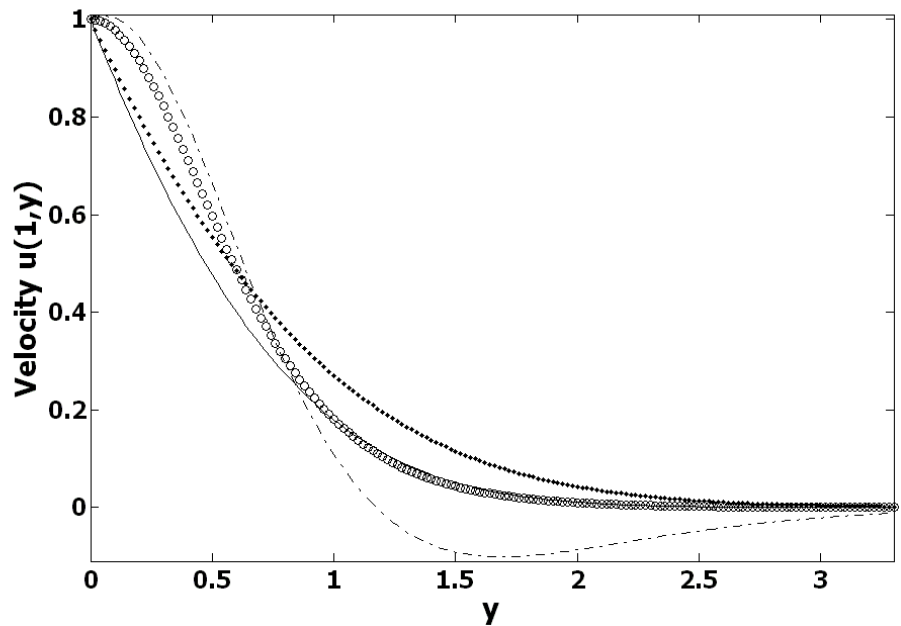

Figure 2. Variation of horizontal-speed/velocity *u* with *y* for present result *K₁=0* (solid curve); *K₁=0.01* (dotted curve); *K₁=0.05* (dotted-dashed curve); and comparison of the results with BS-solution Eq. (24) which is represented here by circle for the Newtonian fluids, when $\alpha_1 = 1.5$.

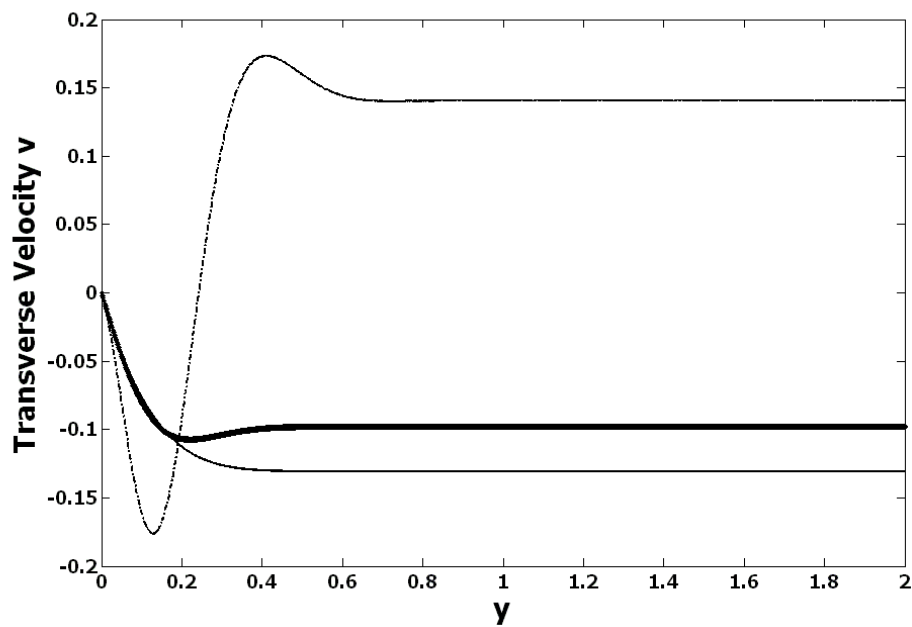

Figure 3. Variation of transverse-speed/velocity *v* with *y* for *K₁=0* (solid curve); *K₁=0.01* (dotted curve) and *K₁=0.02* (dotted-dashed curve)

It is obvious to say that the microrotation $S$ of the aciculate particles in a fluid is zero for the Newtonian fluid since this fluid property occurs in micropolar fluid merely. Therefore $S$ is plotted versus $y$ for non-zero values of $K_1 \in \{0.007, 0.02, 0.05\}$ for a couple of values of $x \in \{0.1, 1.3\}$ in figures 4 & 5. From these figures, it is observed that $S$ depends upon $x$, $y$ and $K_1$. The dependency of $S$ on $y$ is significant in such a way that (i) $S$ increases with increasing $K_1$ for high value of about $x \geq 1$ for all $y$ but it is not so for low value of $x$; (ii) $S$ also increases as $y$ increases in the vicinity of jet along $y$-direction and it decreases as $y$ increases such that $S \to 0$ as $y \to \infty$ which is in accordance with the fact given in (10), analogous to the fluid speed $u$.

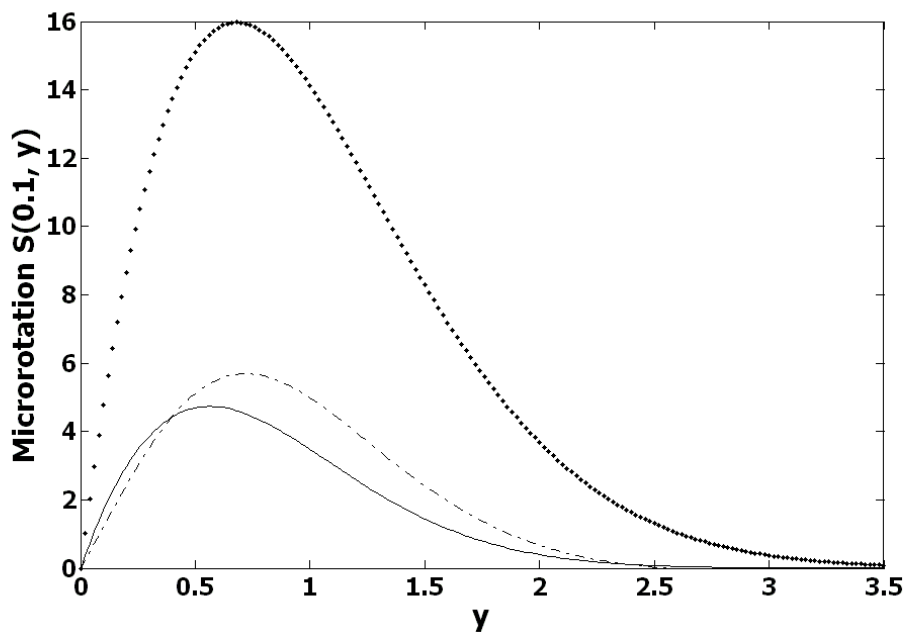

Figure 4. Variation of Microrotation $S$ with $y$ at $x=0.1$ for $K_1=0.007$ (solid curve); $K_1=0.02$ (dotted curve) and $K_1=0.05$ (dotted-dashed curve).

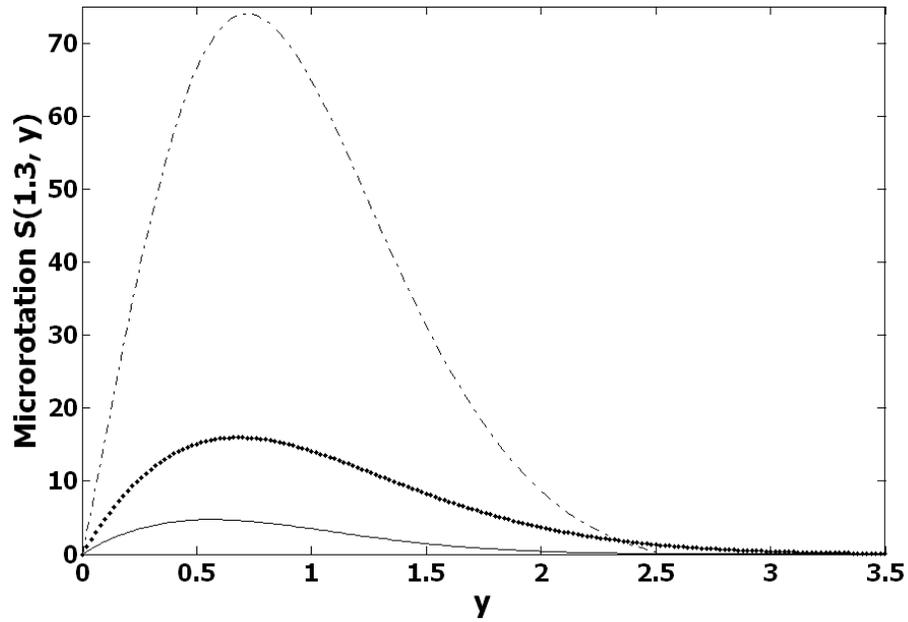

Figure 5. Variation of Microrotation $S$ with $y$ at $x=1.3$ for $K_1=0.007$ (solid curve); $K_1=0.02$ (dotted curve) and $K_1=0.05$ (dotted-dashed curve).

Next, the fluid flux (or discharge) has a dominant role in analyzing fluid dynamics regarding the engineering perspective, it is defined as [12]: $Q = \int_0^\infty u\,dy$. In the present context, it will be deformed in a very simple form as:

$$Q = xf(\infty) \tag{25}$$

by virtue of (13). This Eq. (25) shows that $Q \propto x$ for a fixed value of $K_1$ since $f(\infty)$ varies with $K_1$. Furthermore $Q$ increases as $x$ increases (in downstream too) for both the fluids (Newtonian as well as micropolar) as shown in figure (6), which is in agreement with the existing result (for Newtonian fluid) [12].

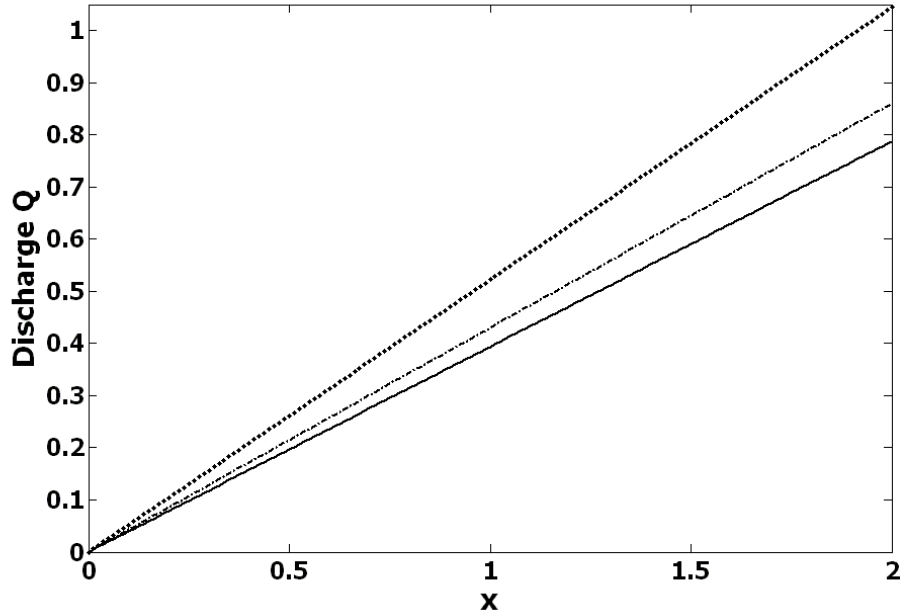

Figure 6. Variation of discharge/fluid-flux $Q$ with $x$ for $K_1=0$ (solid curve); $K_1=0.01$ (dotted curve) and $K_1=0.05$ (dotted-dashed curve).

In our case, Eq. (4) yields two non-zero shear stresses $(\sigma_{12} \ \& \ \sigma_{21})$ as:

$$\sigma_{12} = (1-K_1)\frac{\partial u}{\partial y} + \frac{\partial v}{\partial x} - K_1 S, \tag{26}$$

$$\sigma_{21} = \frac{\partial u}{\partial y} + (1-K_1)\frac{\partial v}{\partial x} + K_1 S, \tag{27}$$

and two non-zero normal stresses $(\sigma_{11} \ \& \ \sigma_{22})$. The normal stresses are independent of x in addition to the fact that $\sigma_{11} = -\sigma_{22}$. Whence, we can conclude that $tr(\sigma_{ij}) = 0$, that implies zero effect of pressure, which is in accordance with the aforementioned assumption made in section 2. The variation of normal stress $\sigma_{11}$ with y for different values of $K_1$ is presented in figure 7. This figure indicates that (i) $\sigma_{11}$ decreases as $y$ increases. It means that the magnitudes of normal stresses are dominant in the vicinity of jet; (ii) $\sigma_{11}$ increases with increasing $K_1$ in the vicinity of jet, for the Newtonian as well as the micropolar fluids. Unlike $\sigma_{11}$ the shear stress, $\sigma_{12}$ increases as $y$ increases for fixed x for a Newtonian fluid while it is not so for a micropolar fluid. For a micropolar fluid, $\sigma_{12} < 0$ and $\sigma_{12}$ decreases initially near jet and then it increases until it converges to zero for far from the jet for all $x$, $K_1$ and for both the fluids (Newtonian and micropolar), as shown in figure 8. The influence of the second non-zero component of shear stress $\sigma_{21}$ is displayed in figure 9. It depicts that (i) $\sigma_{21}$ increases with $y$ for a fixed $x$ for a Newtonian fluid; but (ii) $\sigma_{21}$ increases as $K_1$ increases for all $y$ for a fixed $x$ for the Newtonian as well as micropolar fluids; and (iii) $\sigma_{21} \to 0$ as $y \to \infty$ $\forall x$ and $\forall K_1$ for the Newtonian as well as micropolar fluids. However $\sigma_{21} = \sigma_{12}$ for the Newtonian fluids.

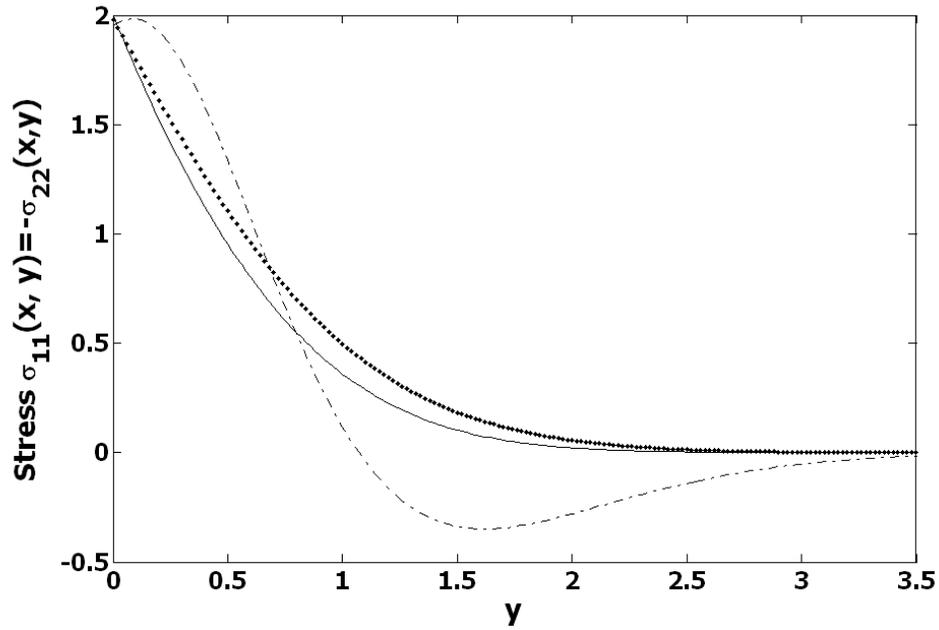

Figure 7. Variation of normal stress $\sigma_{11}$ with $y$ for $K_1=0$ (solid curve); $K_1=0.02$ (dotted curve) and $K_1=0.05$ (dotted-dashed curve).

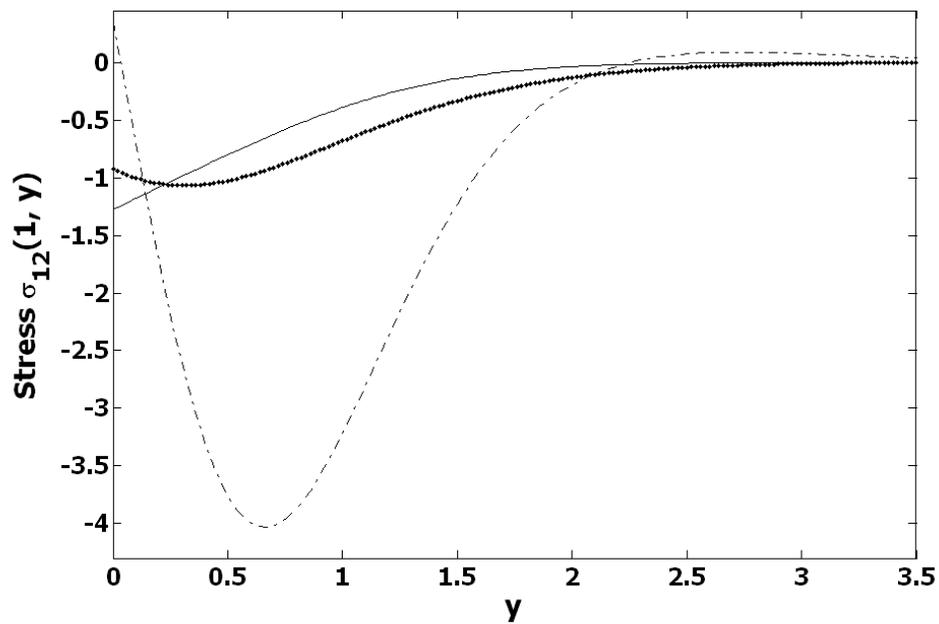

Figure 8. Variation of shear stress $\sigma_{12}$ with *y* at *x=1* for *K₁=0* (solid curve); *K₁=0.02* (dotted curve) and *K₁=0.05* (dotted-dashed curve).

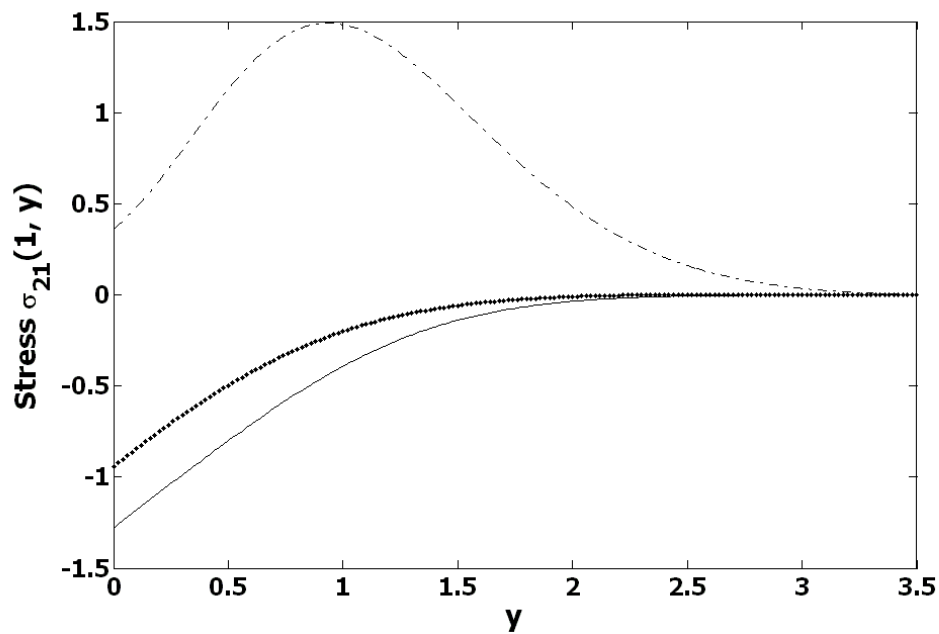

Figure 9. Variation of shear stress $\sigma_{21}$ with *y* at *x=1* for *K₁=0* (solid curve); *K₁=0.02* (dotted curve) and *K₁=0.05* (dotted-dashed curve).

Finally, Eq. (5), for couple stress tensor, yields four non-zero components, which are related in our case as:

$$m_{31} = K_4 m_{13}$$
$$m_{32} = K_4 m_{23}$$
(28)

where $K_4 = \dfrac{\beta}{\gamma} \in [-1,1]$ [22].

According to Eq. (28), the couple stress tensor will be symmetric and skew-symmetric, respectively, if $\beta \mp \gamma = 0$.

The effect of two non-zero components of shear stresses $(m_{13} \ \& \ m_{23})$ are presented graphically here in figures 10 & 11. These figures indicate that $m_{13}$ increases as $K_1$ increases at a fixed $x$ for low value of $y$ and after then it decreases for high value of $y$ such that $m_{13} \to 0$ as $y \to \infty$ for all $K_1$. Unlike $m_{13}$, $m_{23}$ decreases if $y$ increases for all $K_1$ at a fixed x. Consistently, $m_{23} \to 0$ as $y \to \infty$ for all $K_1$ and $x$, analogous to aforementioned physical parameters of fluids.

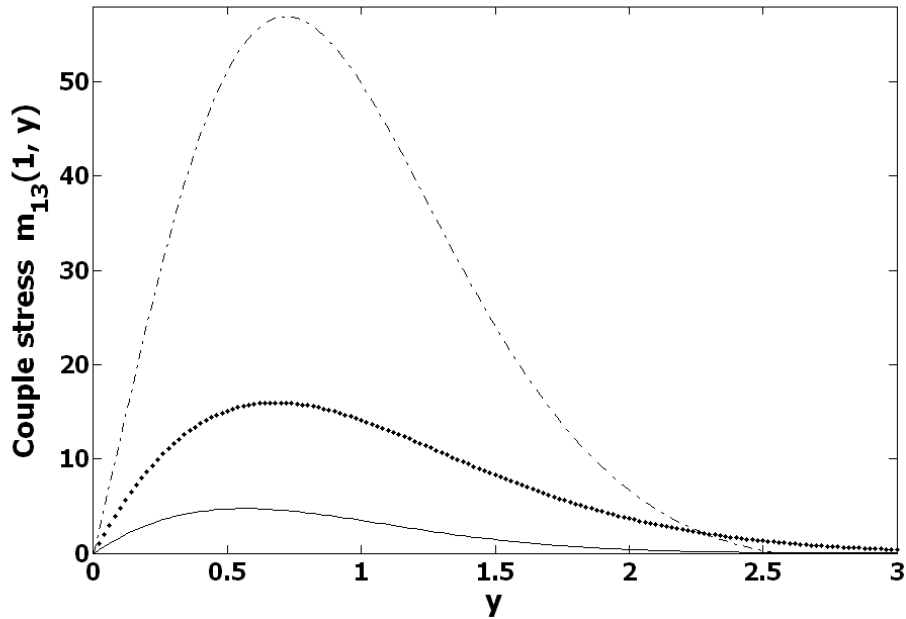

Figure 10. Variation of couple stress $m_{13}$ with $y$ at $x=1$ for *$K_1$=0.007* (solid curve); *$K_1$=0.02* (dotted curve) and *$K_1$=0.05* (dotted-dashed curve).

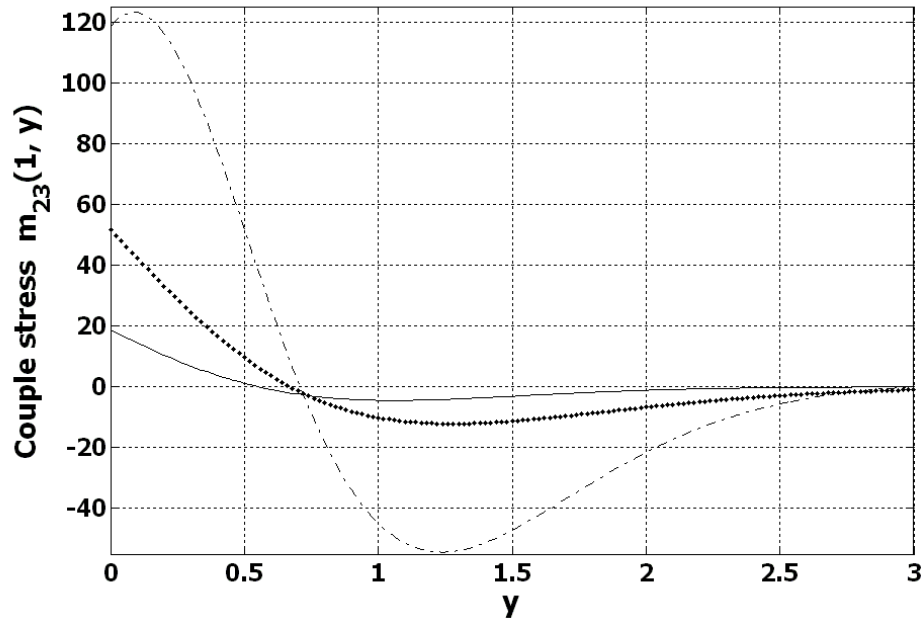

Figure 11. Variation of couple stress $m_{23}$ with $y$ at $x=1$ for $K_1=0.007$ (solid curve); $K_1=0.02$ (dotted curve) and $K_1=0.05$ (dotted-dashed curve).

## 6. Conclusions

In this study, it was formulated the boundary-value-problem (BVP), comprising partial differential equations (PDEs), of steady flow for plane, laminar jet of a micropolar fluid. A new similarity transformation/solution was derived which is valid not only for the Newtonian fluids but also for the micropolar fluids. Obviously, this transformation will be transformed the PDEs into the ordinary differential equations (ODEs). These ODEs were solved numerically by the finite difference method. The obtained results were compared with existing results [1, 12] for the Newtonian fluids. The comparison was favourable. As the aciculate particles in a micropolar fluid can rotate without translation, the micropolarity must have influence on fluid-physical-parameters (e.g. fluid-speed, internal stresses, couple stresses and discharge). This influence was highlighted in the present study. All the fluid-physical-parameters had significant influence on the viscosity coupling parameter $K_1$. For example, if $K_1$ increases, the microrotation, the fluid-flux, the stresses and the couple stresses $(m_{13}\ \&\ m_{31})$ grows in the vicinity of the jet along $y$-direction. It is not amazing to record the observation that all fluid-physical-parameters (FPPs) vanishes far from the jet ($y\rightarrow\infty$) owing to the stationary fluid far from the jet, which may justify our numerical results. In addition to the dependency of FPPs on $K_1$, the horizontal fluid speed, shear stresses and microrotation depend upon $x$ and $y$. In contrast, transverse fluid speed is independent of abscissa $x$ and fluid flux is independent of ordinate $y$ such that fluid flux $Q \propto x$, for a fixed value of $K_1$ and for the micropolar as well as the Newtonian fluids. Moreover, the normal stresses are related mutually as $\sigma_{11} = -\sigma_{22}$; whence we can conclude that $tr(\sigma_{ij}) = 0 \, \forall K_1 \in (0,1)$. Last but not the least, for the micropolar fluids, the stress and the couple stress tensors are non-symmetric but the couple stress tensor will be symmetric and skew-symmetric, respectively, if $\beta \mp \gamma = 0.$